# A statistical significance test for spatio-temporal receptive field estimates obtained using spike-triggered averaging of binary pseudo-random sequences

Murat OKATAN

Istanbul Technical University, Informatics Institute, Istanbul, Türkiye
Department of Artificial Intelligence and Data Engineering, Istanbul Technical University, Türkiye

okatan@itu.edu.tr
ORCID: 0000-0002-0064-6747

**Abstract**

Spatio-temporal receptive fields (STRF) of visual neurons are often estimated using spike-triggered averaging of binary pseudo-random stimulus sequences. The spike train of a visual neuron is recorded simultaneously with the stimulus presentation. The neuron's STRF is estimated by averaging the stimulus frames that coincide with spikes at fixed latencies. Although this is a widely used technique, an analytical method for determining the statistical significance of the estimated value of the STRF pixels seems to be lacking. Such a significance test would be useful for identifying the significant features of the STRF and investigating their relationship with experimental variables. Here, the distribution of the estimated STRF pixel values is derived for given spike trains, under the null hypothesis that spike occurrences and stimulus values are statistically independent. This distribution is then used for computing amplitude thresholds to determine the STRF pixels where the null hypothesis can be rejected at a desired two-tailed significance level. It is also proposed that the size of the receptive field may be inferred from the significant pixels. The application of the proposed method is illustrated on spike trains collected from individual mouse retinal ganglion cells.

## 1. Introduction

Action potential (spike) generation is the mechanism by which neurons rapidly transmit electrical signals across distances that are too long for electrotonic conduction [1].

One approach to explore what aspects of the environment a neuron responds to on the average is to take a snapshot of the environment every time the neuron generates a spike and compute the average of those snapshots. For a visual neuron, the environment may represent distributions of photon energies across a patch of the retina. The restricted extent of the environment to which the neuron responds is captured by the concept of a "receptive field" [2–4]. Receptive fields are thought to be formed by hierarchical spatio-temporal filtering of input signals through cascading cell layers with varying spatial and temporal scales [5]. Since the cascade can only respond to the environment with a finite latency, the averaging of the snapshots can be performed at physiologically relevant latencies to fully capture those aspects of the environment that drives the neuron to generate a spike. The introduction of these latencies into the averaging enables the characterization of spatio-temporal receptive fields (STRF). The averaging of the snapshots of the environment, delayed relative to the spike times, is called spike-triggered averaging (STA) or 'reverse correlation' and is widely used in estimating the STRF [5–18]. When the stimulus patterns of interest are sound frequencies, as in the exploration of the response properties of auditory neurons, the same method is used for constructing spectro-temporal receptive fields, which are also abbreviated as STRF [11]. Receptive fields have also been defined in the spatio-temporal frequency domain, which is the space spanned by spatial and temporal frequencies [19]. Receptive fields may also span more than one sensory modality, such as the bimodal neurons of the midbrain and cortex [20, 21], which may be important for the binding of sensory inputs into a unified percept [22].

Various types of stimuli are used in STRF estimation of visual cells, including finite sets of orthonormal stimuli [14, 15], grating-like stimuli [19, 23], randomly moving single or collections of point sources [17, 24], flashing a small bar of light at various locations [25] and pseudo-random binary sequences, also called m-sequences, white noise or non-Gaussian white noise [13, 16]. Among these stimuli, pseudo-random binary sequences allow for the estimation of the STRF in a relatively much shorter amount of recording time since stimulus presentation and response recording are continuous and not interrupted by inter-stimulus-intervals [26].

In binary pseudo-random visual stimulus sequences, the stimulus consists of black and white pixels that flicker randomly at a fixed frame rate [9, 10, 13, 16, 27, 28]. In computing the STA, the stimulus value of each white pixel is taken as 1, while that of each black pixel is taken as -1.

Depending on whether the visual stimuli consist of lines (1-D) or checkerboard patterns (2-D), STRF estimates become 2-D or 3-D pixel maps, respectively, with time-before-spike as the additional dimension. The structural and functional properties of STRF estimates are examined by extracting certain features, such as the spatial extent (functional cell size) [6, 10, 12, 13–16, 29], spatial symmetry, biphasic time course and orientation [9, 13], time to peak [10, 13, 16, 29] and ratio of the integrated or peak amplitudes in the dark and light areas [12, 21]. While such features are computed from estimated STRF pixel values, an analytical test for determining the statistical significance of these values seems to be lacking. To address this issue, here, the exact distribution of the estimated STRF pixel values is derived for binary pseudo-random sequences and given spike trains, under the null hypothesis that spike occurrences and stimulus values are statistically independent. After this distribution is verified using simulated data, it is used for computing amplitude thresholds to determine the STRF pixels where the null hypothesis can be rejected at a desired two-tailed significance level. The significant pixels are then subjected to a connected component analysis to determine the size of the receptive field. The proposed method is expected to contribute quantitatively to the exploration of the STRF structure of visual and other types of neurons whose STRF can be estimated by spike-triggered averaging of binary pseudo-random stimulus sequences.

## 2. Materials and methods

This section starts by presenting the formula for the estimated value of STRF pixels, which is then used for deriving the probability mass function (pmf) of the estimates under the null hypothesis that the neuron's spiking and the stimulus values are statistically independent. Desired percentiles (e.g., 2.5$^{th}$ and 97.5$^{th}$) of the null distribution are defined as amplitude thresholds for identifying the significant pixels. A method that combines Bonferroni correction of significance values and connected component analysis is proposed for determining the size of the receptive field.

Let $s(x, u)$ denote the stimulus intensity at location $x$ and time $u$ in a binary pseudo-random sequence, where the probability of each binary value is 0.5. Here, $x$ is assumed to represent a one-dimensional discrete spatial coordinate, or pixel address, but it can represent a pixel address in a high dimensional stimulus space without any loss of generality. The value of $s(x, u)$ is taken to be either 1 or -1 [9, 10, 28]. It follows that $2^{-1}(1 + s(x, u)) \sim B(1, 0.5)$, where $B(1, 0.5)$ denotes the Binomial distribution with a success probability of 0.5 and number of trials of 1. The estimated value of the STRF pixel at location $x$ and time $t$ is given by Eq. 1 [6, 9, 18]:

$$h(x, t) = \frac{1}{n} \sum_{i=1}^{n} s(x, \tau_i + t), \tag{1}$$

where, $n$ is the number of spikes fired by the neuron for which the STRF is being estimated, $\tau_i$ is the time of the $i^{th}$ spike, and $t \leq 0$ is a latency between the stimulus values and the spike times. In this way the STRF represents the average stimulus intensity that was observed at time $t$ relative to spike times, at location $x$.

Because the value of $s(x, \tau_i + t)$ is either 1 or –1, $(1 + s(x, \tau_i + t))/2 \sim B(1, p(x, \tau_i + t))$, provided that at most one spike is observed during the presentation of each frame. Under the null hypothesis $H_0$ that the neuron's spiking is statistically independent of the stimulus, $p(x, \tau_i + t) = 0.5$. The probability mass function of the STRF pixel values under $H_0$ will be derived first under the constraint that at most one spike is observed during the presentation of each frame. This constraint will then be relaxed.

## 2.1 Probability mass function of the STRF pixel values under $H_0$

The stimulus values being either 1 or –1, the quantity $S_n = \sum_{i=1}^{n} s(x, \tau_i + t)$ that appears in Eq. 1 is an integer that can take on values in the closed interval $[-n, n]$. Below, the derivation of the pmf of $S_n$ under $H_0$ is handled for the cases of odd and even $n$ separately, before formulating it for any $n$. Again, this derivation is initially subject to the constraint that at most one spike is observed during the presentation of each frame. This constraint is later relaxed. Ultimately, the pmf of $h(x, t)$ is obtained from the pmf of $S_n$, noting that $h(x, t) = n^{-1} S_n$ per Eq. 1.

## 2.2 The pmf of $S_n$ when n is odd

If $n$ is odd, then, $S_n$ cannot be even. Therefore, $Pr(S_n = m | H_0, n) = Pr(S_n = m | n) = 0 = (1 - (-1)^m)$ when $m$ is an even integer in the range $[-n, n]$.

Now, let $m$ denote an odd integer in the range $[-n, n]$. Then $Pr(S_n = m | H_0, n)$ is given by Eq. 2:

$$Pr(S_n = m|H_0, n) = \binom{\frac{m+n}{2}}{n} 0.5^n. \tag{2}$$

Finally, for $m$ an even or odd integer in the closed interval $[-n, n]$, $Pr(S_n = m|H_0, n)$ is obtained from the above results as in Eq.3:

$$Pr(S_n = m|H_0, n) = (1 - (-1)^m)\binom{\frac{m+n}{2}}{n} 0.5^{n+1}. \tag{3}$$

### 2.3 The pmf of $S_n$ when n is even

If $n$ is even, then, $S_n$ cannot be odd. We thus have $Pr(S_n = m|H_0, n) = Pr(S_n = m|n) = 0 = (1 + (-1)^m)$, where $m$ is an odd integer in the range $[-n, n]$.

Now, let $m$ denote an even integer in the range $[-n, n]$. Then $Pr(S_n = m|H_0, n)$ is given by Eq. 2.

Finally, for $m$ an even or odd integer in the closed interval $[-n, n]$, $Pr(S_n = m|H_0, n)$ is obtained from the above results as in Eq.4:

$$Pr(S_n = m|H_0, n) = (1 + (-1)^m)\binom{\frac{m+n}{2}}{n} 0.5^{n+1}. \tag{4}$$

### 2.4 The pmf of $S_n$ when n is odd or even

The above results indicate that an expression for $Pr(S_n = m|H_0, n)$ is obtained for $n$ even or odd by multiplying $(-1)^m$ by $(-1)^n$ in Eq.4, yielding Eq. 5 as the pmf of $S_n$ for $m$ an integer in the closed interval $[-n, n]$:

$$Pr(S_n = m|H_0, n) = (1 + (-1)^{m+n})\binom{\frac{m+n}{2}}{n} 0.5^{n+1}. \tag{5}$$

Substituting $m = S_n$ and denoting the null pmf of $S_n$ by $f_S(S_n|H_0, n)$, the following expression is obtained:

$$f_S(S_n|H_0, n) = (1 + (-1)^{n+S_n})\binom{\frac{n+S_n}{2}}{n} 0.5^{n+1}. \tag{6}$$

Finally, dividing $S_n$ by $n$ per Eq. 1, and denoting the null pmf of $h(x, t)$ by $f_h(h(x,t)|H_0, n)$, the following expression is obtained:

$$f_h(h(x,t)|H_0, n) = f_S(nh(x,t)|H_0, n) = (1 + (-1)^{n(1+h(x,t))})\binom{\frac{n(1+h(x,t))}{2}}{n} 0.5^{n+1}. \tag{7}$$

Because the stimulus values are symmetrical about 0 and are equiprobable, $f_h(h(x,t)|H_0, n)$ is symmetrical about 0.

### 2.5 Allowing for the occurrence of more than one spike within a given frame period

In real spike trains, more than one spike may occur per stimulus frame. Let $\tau_i$ and $\tau_{i+1}$ denote the times of two consecutive spikes that happen to occur during the presentation of the same stimulus frame. Then, the property that $(1 + s(x, \tau_i + t))/2 \sim B(p(x, \tau_i + t))$ would hold for the i[th] spike but not for the i+1[st]. This property states that the stimulus value at position $x$ and latency $t$ that coincides with the i[th] spike is 1 with probability $p(x, \tau_i + t)$ and is -1 with the complement probability. Yet, for the i+1[st] spike, there is no uncertainty regarding the stimulus value that will correspond to the same position and latency. Thus, the computation of the pmf of the STRF pixel values under $H_0$ must take this fact into account when multiple spikes are allowed to occur during the presentation of the same stimulus frame.

### 2.5.1 Partitioning the spikes and their contribution to the STRF pixel values

Let $n = \sum_{j=1}^{J} j n_j$, where $n_j$ is the number of spike groups that consist of $j$ spikes that coincide with the same stimulus frame. In a given data set, the maximum number of spikes that are observed during the presentation of a given stimulus frame can be determined. Here, that number is represented by $J$. For the data analyzed here, the maximum number of spikes fired per frame was five across all retinal ganglion cells examined. So, $J \leq 5$ for the present cells.

It follows that $S_n = \sum_{j=1}^{J} S_{n_j}$, where $-j n_j \leq S_{n_j} \leq j n_j$ are random integers contributed to the STRF by the respective spike partitions. Let $f_{S_1}(S_{n_1}|H_0, n_1)$ denote the pmf of $S_{n_1}$. We have $f_{S_1}(S_{n_1}|H_0, n_1) = f_S(S_{n_1}|H_0, n_1)$ from Eq. 6.

For $S_{n_2}$, note that the occurrence of a spike pair can be treated as a single event in itself. Thus, the $n_2$ such events can be viewed as a new point process, where each point contributes 2 or -2 to $S_{n_2}$. Dividing these quantities by 2, we get the original situation, where each point contributes 1 or -1, this time to $S_{n_2}/2$. Thus, generalizing the foregoing notation, $f_{S_2}(S_{n_2}|H_0, n_2) = f_{S_1}(2^{-1} S_{n_2}|H_0, n_2) = f_S(2^{-1} S_{n_2}|H_0, n_2)$, and more generally, $f_{S_j}(S_{n_j}|H_0, n_j) = f_S(j^{-1} S_{n_j}|H_0, n_j)$, for $1 \leq j \leq J$.

With these probabilities available, since the stimulus frames that contribute to these sums are statistically independent, $Pr\left(\cap_{j=1}^{J} S_{n_j} \middle| H_0, n_{1:J}\right) = \prod_{j=1}^{J} f_S\left(j^{-1} S_{n_j} \middle| H_0, n_j\right)$.

Then, the pmf of $S_n$ under $H_0$ and with multiple spikes per frame allowed, becomes:

$$f_S^J(S_n|H_0, n_{1:J}) = \sum_{m_1=-n_1:2:n_1} \sum_{m_2=-n_2:2:n_2} \cdots \sum_{m_J=-n_J:2:n_J} \mathbb{1}\left(S_n == \sum_{j=1}^{J} j m_j\right) \prod_{j=1}^{J} f_S(m_j|H_0, n_j), \quad (8)$$

where the indicator function $\mathbb{1}(\cdot)$ has a value of 1 if its argument is true, and 0 otherwise. Finally, dividing $S_n$ by $n$ per Eq. 1, and denoting the null pmf of $h(x,t)$ by $f_h^J(h(x,t)|H_0, n_{1:J})$, the following expression is obtained:

$$f_h^J(h(x,t)|H_0, n_{1:J}) = f_S^J(nh(x,t)|H_0, n_{1:J}). \tag{9}$$

**2.6 Positive and negative amplitude thresholds**

It may be desirable to set positive and negative amplitude thresholds to detect STRF pixels where $H_0$ is rejected at level $\alpha$. The $(\alpha/2 \times 100)^{th}$ and $([1-\alpha/2] \times 100)^{th}$ percentiles of $f_h^J(h(x,t)|H_0, n)$ can be used as negative and positive amplitude thresholds for this purpose. Let $\theta^-$ denote the $(\alpha/2 \times 100)^{th}$ percentile (inclusive). For $M$, an integer in the closed interval $[-n, n]$:

$$\theta^- = n^{-1} \min_{-n \leq M \leq n} \left\{ M \middle| Pr(h(x,t) \leq Mn^{-1}|H_0, n_{1:J}) \geq \frac{\alpha}{2} \right\} \tag{10}$$

Here, $Pr(h(x,t) \leq Mn^{-1}|H_0, n_{1:J}) = \sum_{m=-n}^{M} f_h^J(mn^{-1}|H_0, n_{1:J})$ is readily obtained from Eq.9 as in Eq. 11 for $M$ an integer in the closed interval $[-n, n]$:

$$Pr(h(x,t) \leq Mn^{-1}|H_0, n_{1:J}) = \sum_{m=-n}^{M} f_S^J(m|H_0, n_{1:J}) \tag{11}$$

Similarly, $\theta^+$ denoting the $([1-\alpha/2] \times 100)^{th}$ percentile (inclusive), for $M$ an integer in the closed interval $[-n, n]$, $\theta^+$ is given by Eq. 12:

$$\theta^+ = n^{-1} \min_{-n \leq M \leq n} \left\{ M \middle| Pr(h(x,t) \leq Mn^{-1}|H_0, n_{1:J}) \geq 1 - \frac{\alpha}{2} \right\} \tag{12}$$

The computation of $\theta^+$ can be accelerated by using Eq. 13, which computes the upper tail probability instead:

$$\theta^+ = n^{-1} \max_{-n \leq M \leq n} \left\{ M \middle| Pr(h(x,t) \geq Mn^{-1}|H_0, n_{1:J}) \geq \frac{\alpha}{2} \right\} \tag{13}$$

Since $f_S^J(nh(x,t)|H_0, n_{1:J})$ is symmetrical about 0, $\theta^+ = -\theta^-$. Note that, since $h(x,t)$ is quantized, any threshold value in the left-open interval $\Theta^- = (\theta^- - 2n^{-1}, \theta^-]$ or the right-open interval $\Theta^+ = [\theta^+, \theta^+ + 2n^{-1})$ will also select the same set of significant pixels as $\theta^-$ or $\theta^+$, respectively.

With these amplitude thresholds, the null hypothesis $H_0$ that the spiking is statistically independent of the stimulus is rejected at level $\alpha$ at pixels where $h(x,t) < \varepsilon^-$ or $h(x,t) > \varepsilon^+$, with $\varepsilon^- \in \Theta^-$ and $\varepsilon^+ \in \Theta^+$.

## 2.7 Bonferroni correction, connected component analysis and the size of the receptive field

When multiple statistical tests are performed across the STRF, many pixels that are found to be statistically significant may actually be false positives. This can be controlled, although perhaps too conservatively, by using Bonferroni correction [30, 31]. If the STRF consists of $N$ pixels, then, Bonferroni correction is implemented by replacing $\alpha$ with $\alpha/N$ in Eq. 10, Eq. 12 or Eq. 13.

Bonferroni correction may cause a large number of true positive pixels to be missed [30, 31]. An alternative approach to identifying the significant pixels that belong to the STRF is to use a connected component analysis on the significant pixels that are detected without Bonferroni correction [11]. Here, a simple connected component analysis is proposed to illustrate this idea.

First, the STRF is thresholded at $\alpha = 0.05$ with Bonferroni correction to determine whether the lightest pixel is found to be significant. If it is, then, the STRF is thresholded again at $\alpha = 0.05$, this time without Bonferroni correction, to identify the pixels that exceed an $\varepsilon^+$. In the resulting binary image of the significant light pixels, the connected component that contains the lightest pixel is determined using the bwconncomp function of MATLAB. This procedure identifies the ON-subfield of the STRF. If the lightest pixel is not found to be significant at $\alpha = 0.05$ with Bonferroni correction, then, it is inferred that the cell does not have a significant ON-subfield. A similar procedure identifies the OFF-subfield when performed on pixels darker than an $\varepsilon^-$.

Once the ON- and OFF-subfields are identified, the width registered at the time of the lightest or the darkest pixel is used as the subfield size [16, 29]. Alternatively, the size of each subfield may be determined as the difference between the maximum and the minimum $x$ coordinates of the pixels. This difference is denoted here by $\Delta x_{max}$. If the ON- or OFF-subfield is not significant, then, its size is accepted to be 0.

## 2.8 Data

The application of the proposed method is illustrated on spike trains collected from individual mouse retinal ganglion cells [32]. The data were recorded in the isolated retina from laboratory mice (Mus Musculus) using a 61-electrode array [10]. STRFs were mapped by projecting gratings of adjacent thin bars, where each bar flickered black or white according to a pseudo-random binary sequence (16.6 ms frame duration). The bar width was 8.3 μm in the data used here. The STRFs were estimated and visualized using the MATLAB code provided with the data [32]. Results are summarized for the seven retinal ganglion cells that the code selects by default. The default settings of the code estimate and show the STRF with 640 rows and 30 columns, thereby using $N = 640 \times 30 = 19200$ pixels. The number of spikes, $n$, that was used in Eq. 1 was 38845, 4908, 23750, 13184, 24896, 3496, 25717 for cells 1-7, respectively.

## 2.9 Simulation

To verify that Eq. 9 is accurate, Eq. 1 was used to estimate the STRF using a simulated spike train. The stimulus sequence was the one used for stimulating the cells analyzed here [32]. The simulated spike train was a homogeneous Poisson process with a rate of 10 Hz, which was sampled at 10 kHz. The difference between the empirical and theoretical cumulative distribution functions of the estimated pixel values of the STRF was computed. For $0 < \alpha < 1$, the theoretical $100(1 - \alpha)\%$ confidence interval of this difference was constructed at each pixel value as follows:

$$CI(\alpha) = n^{-1}\left[B^{-1}\left(\alpha/2, n, F_h^J(h(x,t)|H_0, n_{1:J})\right), B^{-1}\left(1 - \alpha/2, n, F_h^J(h(x,t)|H_0, n_{1:J})\right)\right], \quad (14)$$

where $B^{-1}\left(b, n, F_h^J(h(x,t)|H_0, n_{1:J})\right)$ is the $(100b)^{th}$ percentile of the Binomial distribution with success probability $F_h^J(h(x,t)|H_0, n_{1:J})$, and $F_h^J(h(x,t)|H_0, n_{1:J})$ is the theoretical cumulative distribution function of $h(x,t)$, obtained from Eq. 9:

$$F_h^J(h(x,t)|H_0, n_{1:J}) = \sum_{v \leq h(x,t)} f_h^J(v|H_0, n_{1:J}). \quad (15)$$

Computations were performed under MATLAB (R2021b) on a Mobile Workstation with 32 GB RAM and Intel® Xeon® E-2276 M CPU @ 2.80 GHz, 6 Cores and 12 Logical Processors.

### 3. Results

### 3.1 Simulation results

The simulated spike train contained a total of $n = 15044$ spikes, where $J = 4$ and $n_1 = 12742$, $n_2 = 1044$, $n_3 = 66$, and $n_4 = 4$. The total number of spikes, $n$, ranged between 3496 for cell 6 to 38845 for cell 1 for the same recording duration as the simulation, which was about 25 minutes.

The estimated STRF pixel values ranged between $-0.0371$ and $0.0368$ in the simulated data. The empirical CDF of the estimates was constructed. The theoretical CDF of the same pixel values was computed (Eq. 15). The difference between these two functions is shown in Fig. 1, along with its 95% and 99% confidence intervals (Eq. 14). It is seen that the difference between the functions is within acceptable limits, demonstrating the accuracy of the derived theoretical CDF.

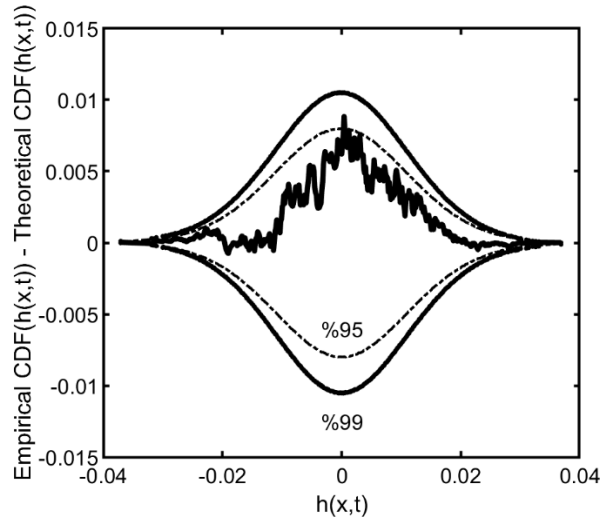

**Fig. 1** Difference between the empirical and theoretical cumulative distribution functions for the simulated data. The difference lies completely within the 99% confidence interval and almost completely within the 95% confidence interval.

**3.2 Thresholding the STRF estimates in real data**

Fig. 2 illustrates STRF estimates for the retinal ganglion cells 1, 4 and 7, where the left column graphs show $h(x,t)$, and the middle and the right column graphs show the significant pixels at level $\alpha = 0.05$ with or without Bonferroni correction, respectively. The darkest and the lightest pixels are detected as significant in the STRFs of cells 1 and 4 at level $\alpha = 0.05$ with Bonferroni correction. In the STRF of cell 7, however, they are detected only without Bonferroni correction (bottom right graph in Fig. 2)[1]. Note that the location of the lightest pixel (indicated by a white circle) does not seem to be related to the STRF of cell 7.

Fig. 3 shows the OFF- and ON-subfields of the STRF in detail for cell 1. While the ON-subfield consists mostly of a large contiguous blob, the OFF-subfield is fragmented into several smaller blobs.

The computation of the thresholds took 1.8, 2.2, 4.2, 13.9, 18.3, and 57.5 minutes for cells 7, 4, 5, 1, 2, and 6, respectively, while it took 12.2 hours for cell 3. The duration increases with the number of terms in Eq. 8.

---

[1] Revised sentence (see Acknowledgment).

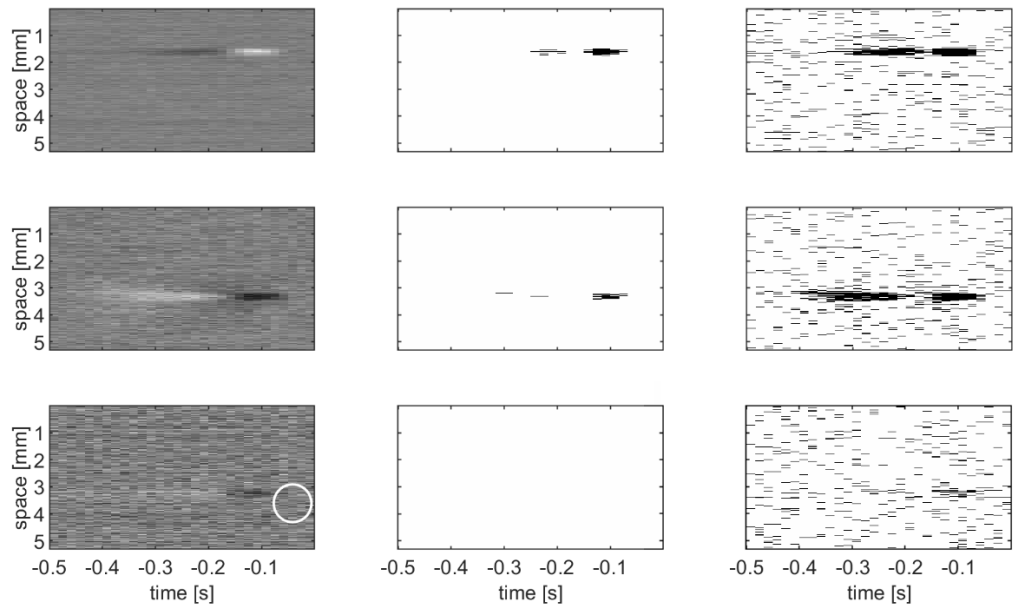

**Fig. 2** STRFs of three retinal ganglion cells. Left panels show the plot of Eq. 1 for cells 1 (top), 4 (middle) and 7 (bottom). Pixel values range from -1 (black) to 1 (white). The location of the lightest pixel is indicated by the center of a white circle for cell 7. The STRFs are thresholded using Eq. 10 and Eq. 12. Significant pixels are shown in black against a white background for $\alpha = 0.05$ with (middle column) and without (right column) Bonferroni correction

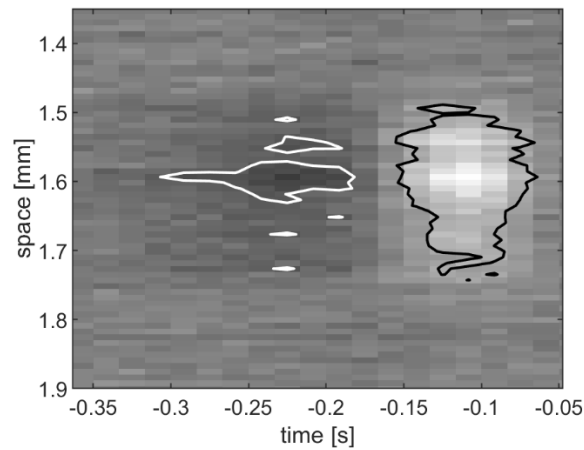

**Fig. 3** Detailed view of the STRF of cell 1. Contour lines show the $\theta^-$ and $\theta^+$ thresholds in white and black, respectively, for $\alpha = 0.05$ with Bonferroni correction

### 3.3 Connected component analysis results and the size of the receptive field

Figure 4 shows the connected components that contain the pixels where the minimum and the maximum STRF values are observed for cell 1 (obtained from the top right panel of Fig. 2). For the OFF-subfield, $\Delta x_{max}$ is measured as 273.9 µm, while the width at the darkest pixel is 240.7 µm. For the ON-subfield, $\Delta x_{max}$ and the width at the lightest pixel are measured as 265.6 µm and 257.3 µm, respectively. Receptive field sizes of all analyzed cells are listed in Table 1.

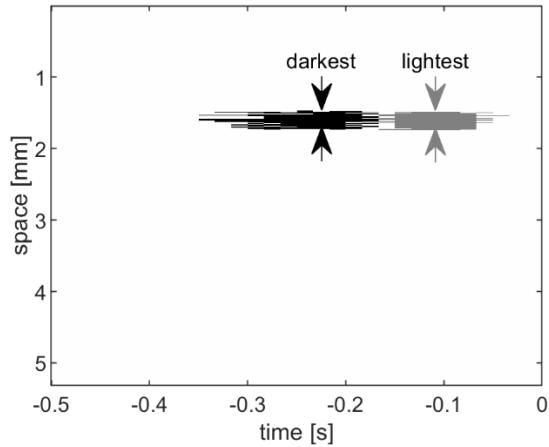

**Fig. 4** Connected components and the size of the receptive field. Connected components show the OFF- (black) and ON-subfields (gray) of the STRF. Arrows indicate the width of the receptive field at the times of the darkest and the lightest pixels

Table 1. Receptive field sizes of the analyzed cells (µm)

| Cell number | OFF-subfield | | ON-subfield | |
|---|---|---|---|---|
| | At the darkest pixel | $\Delta x_{max}$ | At the lightest pixel | $\Delta x_{max}$ |
| 1 | 240.7 | 273.9 | 257.3[a] | 265.6[a] |
| 2 | 0 | 0 | 132.8[a] | 141.1[a] |
| 3 | 224.1[a] | 240.7[a] | 91.3 | 107.9 |
| 4 | 315.4[a] | 365.2[a] | 315.4 | 332.0 |
| 5 | 199.2 | 224.1 | 390.1[a] | 398.4[a] |
| 6 | 132.8 | 141.1 | 174.3[a] | 174.3[a] |
| 7 | 0[a] | 0[a] | 0 | 0 |

[a] These values may be taken as the size of the overall receptive field (see Discussion).

### 4. Discussion

Estimation of the STRF using STA provides a relatively straightforward method for characterizing the average stimulus features to which a neuron responds. The STRFs in Figs. 3 and 4 clearly show the spatio-temporal ON- and OFF-subfields of the retinal ganglion cell responses. However, as the thresholded STRF plots show, some or all of the pixels of these response fields fail to reach statistical significance for some cells. Since certain features of the estimated STRFs, such as the spatial extent (functional cell size) [6, 10, 12], spatial symmetry, biphasic time course and orientation [9], time to peak [10] and ratio of the integrated amplitudes in the dark and light areas [12] are used to quantify neural function; such analyses would be less affected by noise if these features are extracted from STRFs that pass the statistical significance test at a sufficiently large number of relevant pixels.

Our simulation result demonstrated that the theoretical distribution deviated from the empirical distribution by an error that was fully contained within the 99% confidence interval of the error (Fig. 1). This was a realistic simulation, where more than 15,000 spikes were used, and $J$ was 4.

The analysis of real data took a couple of minutes to about 12 hours to complete for different cells depending on how many different combinations of $S_{n_j}$'s could be added up to obtain $S_n$ ($S_n = \sum_{j=1}^{J} S_{n_j}$, see Section 2.2.1).

The main reason for the long computation times is the large number of frames in which more than one spike is observed. For cell 3, $n_{1:4}$ were 10193, 5984, 523 and 5, respectively.

Figure 2 shows that thresholding the STRF estimate without Bonferroni correction detects several pixels that are scattered throughout the graphs (right panels)[2]. Although Bonferroni correction clears them out to a large extent, this may lead to an increased rate of Type I error [30].

In order to combat this, the results of thresholding with and without significance correction are combined to identify the STRF as the connected components that contain the lightest and the darkest pixels, provided that the latter are significant at level $\alpha = 0.05$ with Bonferroni correction (Fig. 4). The size of the receptive field is then readily computed from the spatial coordinates of the pixels belonging to the identified connected components (Table 1, Fig. 4).

The size of the receptive field is an important feature of the STRF [6, 10, 12, 13, 15, 16, 29]. It is proposed here that an estimate of the receptive field size may be readily determined from the significant pixels. This is illustrated here by determining the size of the ON- and OFF-subfields of the STRF (Fig. 4). Specifically, quantities such as the maximum width ($\Delta x_{max}$) or the width registered at the darkest or the lightest pixel may be used as the size of the ON- and OFF- subfields of the STRF, respectively.

The cells analyzed here are termed ON-cells or OFF-cells depending on whether the rightmost subfield of their STRF is an ON or OFF response, respectively [10]. According to this definition, cells 1, 2, 5 and 6 are ON-cells. Thus, for these cells, the size of the ON-subfield may be used as the overall receptive field size, while for the remaining cells, the size of the OFF-subfield may be used, to compare the present results with previous ones that were obtained for these cells.

Overall, the present results suggest that the exact statistical test formulated here may be used to quantitatively probe the functional properties of STRFs and their subfields.

---

[2] Revised sentence (see Acknowledgment).

## 5. Conclusions

An exact statistical test is developed for determining significant pixels for desired two-tailed significance levels in STRF estimates that are obtained using spike-triggered averaging of binary pseudo-random stimulus sequences for given spike trains. A connected component analysis is proposed to identify significant pixels that belong to the STRF. It is proposed that the width of the rightmost connected component at the time of the lightest or the darkest pixel may be used as the size of the receptive field. The test can also be used with Bonferroni corrected significance levels to examine the fine structure of the STRF subfields.

**Acknowledgment** The original version of this article was first published in Okatan, M. A statistical significance test for spatio-temporal receptive field estimates obtained using spike-triggered averaging of binary pseudo-random sequences. SIViP 17, 3759–3766 (2023). https://doi.org/10.1007/s11760-023-02603-1, by Springer Nature. The current manuscript corrects an error in the lower right panel of Fig. 2 of the original version. The sentences indicated by footnotes 1 and 2, which refer to Fig. 2, are revised accordingly.

**Author contributions** MO identified the problem, developed the method, performed the analyses and wrote the manuscript.

**Funding** The author did not receive support from any organization for the submitted work.

**Availability of data and materials** The datasets analyzed during the current study are available in the CRCNS.ORG repository, http://dx. doi.org/10.6080/K0RF5RZT [32].

## Declarations

**Conflict of interest** No, I declare that the authors have no competing interests, or other interests that might be perceived to influence the results and/or discussion reported in this paper. The author has no relevant financial or non-financial interests to disclose.

**Ethical approval** Not applicable (No ethical approval was needed for the present study since the data were downloaded as computer files from http://dx.doi.org/10.6080/K0RF5RZT [32]).